\documentclass{article}

\usepackage{PRIMEarxiv}

\usepackage[utf8]{inputenc} 
\usepackage[T1]{fontenc}    
\usepackage{hyperref}       
\usepackage{url}            
\usepackage{booktabs}       
\usepackage{amsfonts}       
\usepackage{nicefrac}       
\usepackage{microtype}      
\usepackage{lipsum}
\usepackage{fancyhdr}       
\usepackage{graphicx}       
\graphicspath{{media/}}     

\usepackage{xspace} 
\newcommand{\ie}{\textit{i.e.}\xspace} 
\newcommand{\eg}{\textit{e.g.}\xspace} 

\usepackage{tcolorbox} 
\usepackage[ruled,vlined]{algorithm2e} 
\usepackage{natbib}
\usepackage{multirow}
\usepackage{amsmath}  

\title{Principal Context-aware Diffusion Guided Data Augmentation for Fault
Localization
}

\author{
  ShihaoFu  \\
  School of Big Data and Software Engineering \\
  Chongqing University \\
  Chongqing\\
  \texttt{fushihaocqu@163.com} \\
   \And
  YanLei \\
  School of Big Data and Software Engineering \\
  Chongqing University \\
  Chongqing\\
}

\newcommand{\appname}{PCD-DAug }

\begin{document}
\maketitle

\begin{abstract}
Test cases are indispensable for conducting effective fault localization (FL). 
However, test cases in practice are severely class imbalanced, 
\ie the number of failing test cases (\ie minority class) is much less than that of passing ones (\ie majority class).
The severe class imbalance between failing and passing test cases have hindered the FL effectiveness. 

To address this issue, we propose \appname: a \underline{\textbf{P}}rincipal \underline{\textbf{C}}ontext-aware \underline{\textbf{D}}iffusion guided \underline{\textbf{D}}ata \textbf{\underline{Aug}}mentation approach that generate synthesized failing test cases for improving FL.
\appname first combines program slicing with principal component analysis to construct a principal context that shows how a set of statements influences the faulty output via statistical program dependencies.
Then, \appname devises a conditional diffusion model to learn from principle contexts for generating synthesized failing test cases and acquiring a class balanced dataset for FL.
We conducted large-scale experiments on six state-of-the-art FL approaches and compare \appname with six data augmentation baselines. The results show that \appname significantly improves FL effectiveness, \eg achieving average improvements of 383.83\%, 227.08\%, and 224.19\% in six FL approaches under the metrics Top-1, Top-3, and Top-5, respectively. 
\end{abstract}


\section{Introduction}\label{sec:introduction}

With the rapid development of large-scale software systems, effective fault localization (FL) methods have become crucial. Over the years, numerous FL approaches (\eg~\cite{jones2004fault,jones2005empirical,li2019deepfl,li2021fault,naish2011model,sohn2017fluccs,wen2019historical,wong2012software,zhang2019cnn,lei2023mitigating,hu2024deep,hu2023light,xie2022universal}) have been proposed to identify faulty statements in programs. These approaches typically fall into two categories: spectrum-based fault localization (SFL)~\cite{naish2011model,abreu2007accuracy} and deep learning-based fault localization (DLFL)~\cite{zhang2019cnn,zhang2021study,zheng2016fault}. Both approaches rely on the execution information of test cases, \eg coverage information denoted as a statement \textit{executed} or \textit{not executed}, and test results represented as a \textit{passing} or \textit{failing} result. 
Based on the execution information of test cases, FL approaches apply suspiciousness evaluation algorithms, such as correlation coefficients for SFL or neural networks for DLFL, to rank program statements based on their likelihood of being faulty~\cite{kochhar2016practitioners, tan2017codeflaws}. 

Thus, test cases are indispensable for conducting effective fault localization.
FL approaches classify test cases into two classes: passing test cases and failing ones, to analyze their distinct behaviors and pinpoint the locations of a fault.
However, a significant challenge is the class imbalance between passing and failing test cases, 
where failing test cases are often severely less in number. 
This class imbalance can introduce bias into the suspiciousness evaluation~\cite{sun2009classification, he2009learning}, and prior research~\cite{gong2012effects} has shown that a more balanced dataset can improve the effectiveness of FL. 

Recent research has focused on developing data augmentation approaches for FL that generate balanced test suites to enhance the effectiveness of FL.
As modern software systems grow increasingly complex, the dimensionality of the execution information (\eg coverage information regarding the size of a program) of test cases becomes exceedingly high,
these data augmentation approaches usually perform dimensionality reduction before data generation.
Thus,
a typical workflow involves two parts:
dimensionality reduction and data generation,
Specifically, 
it first applies dimensionality reduction techniques to the test cases,
followed by data generation approaches to generate new failing test cases until the dataset becomes balanced, where the number of failing test cases is equal to that of passing ones.

In the dimensionality reduction phase, current approaches~\cite{lei2023mitigating,hu2024deep,hu2023light,xie2022universal} uses either program slicing~\cite{weiser2009program} or linear dimensionality reduction techniques (\eg linear discriminant analysis~\cite{xanthopoulos2013linear} and principal component analysis~\cite{abdi2010principal}). 
Despite the promising FL results delivered by these existing approaches, 
they are still limited. 
Program slicing focuses on the context of a program based on semantic properties, 
while linear dimensionality reduction simplifies the dataset based on statistical properties. 
These approaches, 
when used separately,  
can fail to capture both the semantic context of a program and the global statistical information.

In the data generation phase, current approaches can be roughly categorized into traditional transformation approaches and deep learning-based ones. Traditional transformation approaches directly modify failing test cases to generate new ones, \eg Lamont ~\cite{hu2023light} applied SMOTE, and PRAM ~\cite{hu2024deep} used a Mixup approach\cite{zhang2017mixup} inspired by image augmentation techniques. While effective, these approaches struggle to capture deeper features in the data. To further capture deep features, deep learning-based approaches, \eg generative adversarial networks (GANs\cite{goodfellow2020generative}) and conditional variational autoencoder (CVAE\cite{kim2021conditional}), train models that learn the characteristics of both passing and failing test cases to generate new failing test cases. \cite{xie2022universal,lei2023mitigating} have demonstrated the effectiveness of using CVAE and GAN for data augmentation in FL, particularly for DLFL. However, these approaches share a common limitation: both GANs and CVAE consist of two components, \ie GANs with a generator and discriminator, and CVAE with an encoder and decoder. These components are interdependent during training, and the potential for capability mismatches between them can result in unstable sample quality\cite{saxena2021generative}.

To address these issues, we propose \appname: a \underline{\textbf{P}}rincipal \underline{\textbf{C}}ontext-aware \underline{\textbf{D}}iffusion guided \underline{\textbf{D}}ata \textbf{\underline{Aug}}mentation approach that generate synthesized failing test cases for improving FL.
The basic idea of \appname is to combines program semantic properties (\ie program slicing) with statistical data properties (\ie principal component analysis) to construct a dimensionality reduced context, and use component-independent deep learning networks (\ie diffusion model) to learn from the context to generate minority class data (\ie failing test cases) for improving FL.

For acquiring a dimensionality reduced context,
\appname
uses dynamic program slicing~\cite{agrawal1990dynamic} to capture the program semantic properties via program dependencies showing how a set of statements influences the faulty output, and leverages a revised principal component analysis (PCA)~\cite{song2010feature} to extract global statistical data properties from the coverage data; then embodies the two properties to define a principal context.
For generating synthesized failing test cases, 
\appname uses the conditional diffusion model~\cite{ho2022classifier} to learn the principal context through interdependent components. Unlike GAN and CVAE requiring the training of two interdependent components, diffusion models\cite{ho2020denoising} consist of two processes: a deterministic forward process and a trainable reverse process. The forward process, which progressively adds noise to degrade the original data, is defined by a mathematical formula and requires no training. The training of a model training focuses solely on the reverse process, which learns to progressively denoise the data to recover failing test cases.

It not only avoids the capability mismatch common in GANs and CVAE but also simplifies the training process. By training only the reverse denoising process, we can achieve more efficient and stable generation of failing test cases, ultimately improving FL effectiveness.

To evaluate \appname, we conducted large-scale experiments on 262 versions across five benchmarks. We applied \appname to six state-of-the-art FL approaches and compared it with six data augmentation approaches. The experimental results show that \appname significantly improves the effectiveness of all six FL approaches and outperforms the six data augmentation approaches.
For example, compared to the six state-of-the-art FL methods, \appname improves FL effectiveness by an average of 383.83\%, 227.08\%, and 224.19\% on the Top-1, Top-3, and Top-5 metris, respectively; 
compared to SOTA of data augmentation approaches, \appname achieves an improvement of 34.51\%, 0.56\%, and 3.40\% on the same metrics.

The main contributions of this paper can be summarized as follows:

\begin{itemize}
    \item We propose \appname: a principal context-aware diffusion guided data augmentation approach integrating principal context with a diffusion model, generating synthesized failing test cases to acquire a class balanced dataset for improving FL.

    \item We devise a principal context combining program semantic properties (\ie program slicing) with statistical data properties(\ie revised PCA), guiding the data synthesis process within a diffusion model framework.
 
    \item We conduct comprehensive experiments involving six state-of-the-art FL techniques, alongside six data augmentation approaches. Our results show that \appname significantly improves the FL effectiveness.

    \item We open source the replication package online\footnote{\url{https://github.com/sh10f/PCD-DAug}}, including the all source codes.
\end{itemize}
	
The rest of this paper is structured as follows.
	Section~\ref{background} introduces background information.
	Section~\ref{approach} presents our approach \appname.
	Section~\ref{experiments} and Section~\ref{discussion} show the experimental
	results and discussion.
	Section~\ref{conclusion} draws the conclusion.

\section{Background}\label{background}

\subsection{Diffusion Model}
Diffusion models are a type of generative model applied in tasks such as image generation and text-to-image synthesis. In recent years, diffusion models have gained significant attention from researchers due to their impressive performance in both text-to-image and text-to-video generation tasks. A diffusion model consists of two main components: the forward process, also known as the diffusion process, and the reverse process. In the forward process, Gaussian noise is gradually added at each time step to transform the data into a fully noisy result. The reverse process, in turn, works by predicting and removing the Gaussian noise step by step, ultimately recovering the original sample. The detailed processes are as follows:

\textbf{Forward Process}. 
In the forward process, an original data sample $\mathbf{x}_0$ undergoes a series of transformations where Gaussian noise is added progressively at each time step. This process is modeled as a Markov chain, and can be described mathematically as follows:
\begin{equation} \label{eq2}
    \begin{aligned}
        q(\mathbf{x}_{1:T}|\mathbf{x}_0) &:=\prod_{t=1}^T q(\mathbf{x}_t|\mathbf{x}_{t-1}) \\
        q(\mathbf{x}_t|\mathbf{x}_{t-1}) &:= \mathcal{N}(x_t;\sqrt{1-\beta_t}\mathbf{x}_{t-1}, \beta_t\mathbf{I})
    \end{aligned}
\end{equation}
where $x_t$ represents the data at time step 
t in the forward process, and $x_0$ is the original sample. $\beta_t$ is the variance schedule, which controls the amount of noise added at each step. The variance can either be learned through reparameterization \cite{kingma2013auto} or kept as a constant parameter\cite{ho2020denoising}.

As Gaussian noise is continuously added over time, the original sample \( x_0 \) is eventually transformed into an indistinguishable noisy version, represented by \( x_T \), which approximates a sample from a standard Gaussian distribution. The forward process transforms the data distribution into a noise distribution. This transformation can also be described mathematically as:
\begin{equation}
x_t=\sqrt{\bar{\alpha}_t} x_0+\sqrt{1-\bar{\alpha}_t} \epsilon
\end{equation}
\noindent where $\alpha_t=1-\beta_t$ and $\bar{\alpha}_t=\prod^t_{s=1}\alpha_s$. $\epsilon$ is the Gaussian-distributed noise.

\textbf{Reverse Process}. The reverse process is designed to recover the original data sample $\mathbf{x}_0$ from the fully noisy data $\mathbf{x}_T \sim \mathcal{N}(0, \mathbf{I})$. This reverse transformation is achieved through a step-by-step denoising procedure, modeled by a Markov chain. The reverse process can be expressed as:

\begin{equation}
p_{\theta}(\mathbf{x}_{0:T}) = p(\mathbf{x}_T) \prod_{t=1}^T p_{\theta}(\mathbf{x}_{t-1}|\mathbf{x}_t)
\end{equation}

where $p_{\theta}(\mathbf{x}_{t-1}|\mathbf{x}_t)$ represents the conditional probability of transforming $\mathbf{x}_t$ into $\mathbf{x}_{t-1}$, which is parameterized as a Gaussian distribution:

\begin{equation}
p_{\theta}(\mathbf{x}_{t-1}|\mathbf{x}_t) = \mathcal{N}(\mathbf{x}_{t-1}; \mu_{\theta}(\mathbf{x}_t, t), \Sigma_{\theta}(\mathbf{x}_t, t))
\end{equation}

Where, $\mu_{\theta}(\mathbf{x}_t, t)$ and $\Sigma_{\theta}(\mathbf{x}_t, t)$ are the mean and variance predicted by the model. In most cases, the variance $\Sigma_{\theta}(\mathbf{x}_t, t)$ is set to a constant, and can be defined as:

\begin{equation}
\Sigma_{\theta}(\mathbf{x}_t, t) = \sigma_t^2 \mathbf{I}, \quad \sigma_t^2 = \frac{1 - \bar{\alpha}_{t-1}}{1 - \bar{\alpha}_t} \beta_t
\end{equation}

As for the mean $\mu_{\theta}(\mathbf{x}_t, t)$, it is computed by removing the noise predicted by the model $\epsilon_{\theta}(\mathbf{x}_t, t)$, and is given by:

\begin{equation}
\mu_{\theta}(\mathbf{x}_t, t) = \frac{1}{\sqrt{\alpha_t}} \left( \mathbf{x}_t - \frac{\beta_t}{\sqrt{1 - \bar{\alpha}_t}} \epsilon_{\theta}(\mathbf{x}_t, t) \right)
\end{equation}

In this formulation, $\epsilon_{\theta}(\mathbf{x}_t, t)$ represents the noise component predicted by the model, which is usually parameterized by a neural network. A commonly used architecture for $\epsilon_{\theta}$ is the U-Net~\cite{ronneberger2015u} or Transformer\cite{vaswani2017attention}, which allows efficient denoising at each time step.

By iteratively applying this reverse process, the diffusion model progressively removes the Gaussian noise from $\mathbf{x}_T$, ultimately reconstructing an approximation of the original data $\mathbf{x}_0$.

\textbf{Optimization}. The core objective of the diffusion model is to minimize the difference between the noise added during the forward process and the noise predicted during the reverse process. The diffusion model seeks to align the posterior distribution from the forward process with the prior distribution in the reverse process. This alignment is typically achieved by minimizing the Kullback-Leibler (KL) divergence between the two distributions. For simplification, the objective function can be expressed as a mean squared error (MSE) loss between the true noise and the predicted noise:

\begin{equation}
L_{\text{simple}} = \mathbb{E}_{t, x_0, \epsilon} \left[ \left\| \epsilon - \epsilon_\theta \left( \sqrt{\bar{\alpha}_t} x_0 + \sqrt{1 - \bar{\alpha}_t} \epsilon, t \right) \right\|^2 \right]
\end{equation}

Here, \( \epsilon \) represents the Gaussian noise added to the original data sample \( x_0 \) during the forward process, and \( \epsilon_\theta \) is the noise predicted by the model at step \( t \). The goal of training is to minimize the distance between the added noise and the model’s predicted noise, thereby learning an effective denoising function.

\subsection{Program Slice}

Program slicing is a decomposition technique used to extract the parts of a program that directly or indirectly influence the values computed at a specific program point, referred to as the slicing criterion\cite{lei2012effective, xu2005brief}. A slicing criterion typically consists of a pair $\langle p, V \rangle$, where $p$ is a program location, and $V$ is a set of variables of interest. The subset of the program that affects the values of these variables at $p$ is known as the program slice.

This technique analyzes both control dependencies and data dependencies to determine which parts of the program impact the specified point. By isolating these dependencies, program slicing can aid in identifying the source of program failures, providing a more precise, context-aware view for debugging.

Program slicing can be categorized into static slicing and dynamic slicing\cite{agrawal1990dynamic}, depending on whether the specific input of the program is considered in the slicing process.

\textbf{Static program slicing} does not take specific program inputs into account. It analyzes all possible execution paths based solely on the program’s structure to identify statements that may influence the value of a particular variable. Static slices include all potential paths and are useful for providing a comprehensive analysis of the program's control and data flows, helping developers understand the overall behavior of the program.

\textbf{Dynamic program slicing} is an important technique for debugging, as it includes only the statements along the execution path that affect the value of a variable at a specific program point for a given input. The slicing criterion in dynamic slicing is extended to a triplet $\langle p, V, I \rangle$, where $I$ represents the set of inputs. Dynamic slicing can provide more precise slices by focusing on relevant execution paths, but at the cost of requiring actual program runs.

By comparing the two approaches, dynamic slicing offers more refined results, especially when debugging failures under specific test cases.

\section{Approach}\label{approach}

\begin{figure}[htbp]
\centerline{\includegraphics[scale=0.35]{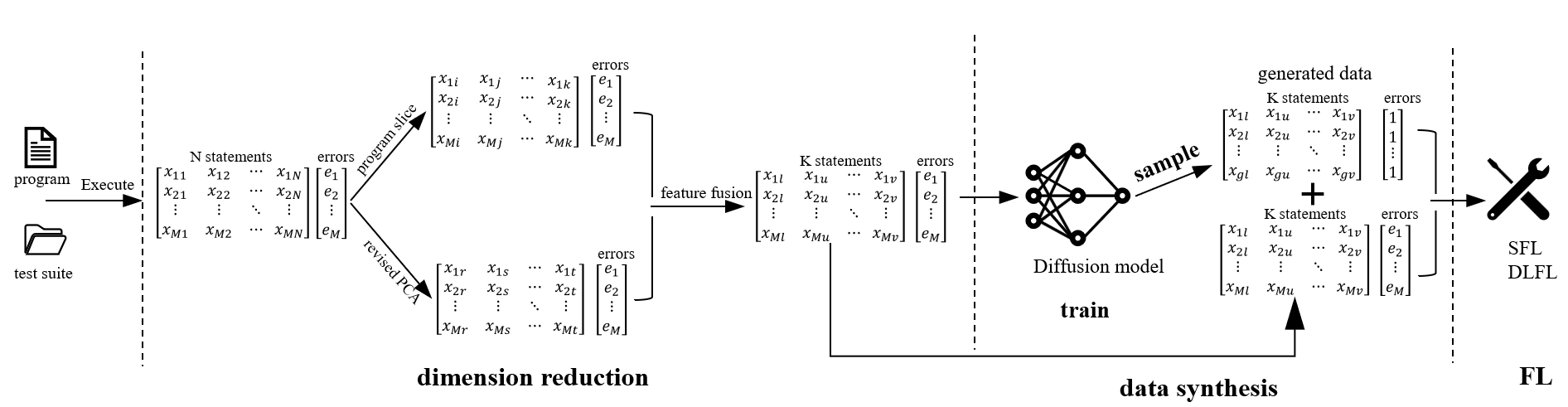}}
    \caption{Architecture of \appname.}
    \label{architecture}
\end{figure}

\begin{figure}[htbp]
\centerline{\includegraphics[scale=0.38]{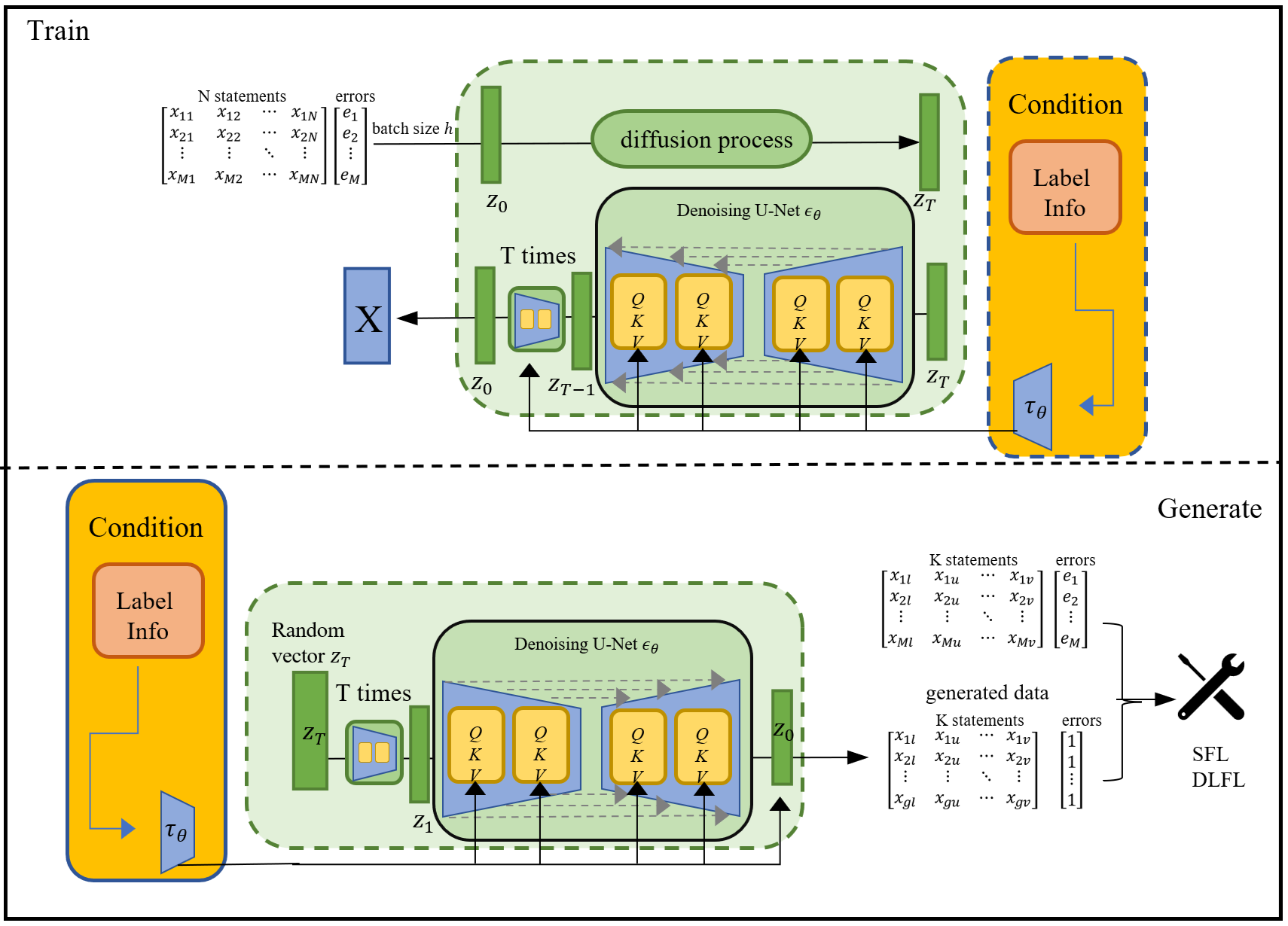}}
    \caption{Data synthesis stage of \appname.}
    \label{synthesis}
\end{figure}

This section will introduce our approach \appname: 
a Context-Aware and PCA-enhanced Diffusion Model for data augmentation in fault localization. As shown in \figurename{}~\ref{architecture}, \appname operates in three main stages:

\appname first applies dynamic program slicing to capture the fault semantic context based on the program's structure and data dependencies. Next, by using a revised PCA on the raw data, we extract the statistical context from a statistical analysis perspective. These two contexts are then merged and fed into the training process of the diffusion model, which requires training only the reverse denoising process.

Finally, the trained context-aware diffusion model is used to generate new failing test cases, iteratively synthesizing data until a class-balanced dataset is achieved, where the number of failing test cases matches the passing ones. This class balance significantly improves the performance of fault localization.

\subsection{Fault Semantic Context Construction}

The fault semantic context refers to the subset of statements whose execution leads to failing outputs. We employ dynamic program slicing to construct this fault semantic context, as it relies on specific program inputs, aligning with the generation process of the raw data (i.e., the coverage matrix and error vector) in FL. The raw data is derived from runtime information collected when executing the test suite on the program. Furthermore, numerous studies\cite{lei2023mitigating, hu2024deep} have shown that dynamic program slicing enhances the effectiveness of FL techniques.

To construct the fault semantic context, we define the dynamic program slicing criterion \textit{ScContext} as:
\begin{equation}\label{slice}
ScContext = \left ( outputStm,outputVar,inputTest \right ) 
\end{equation} 

where \(outputStm\) represents a point of interest in the program, typically a specific statement. \(outputVar\) refers to a set of variables used at \(outputStm\). \(InputVar\) represents the input to the failing test cases. In previous works \cite{lei2023mitigating} and \cite{hu2024deep}, the failing test case with the fewest executed statements was selected for dynamic program slicing to build the context. This approach effectively reduces data dimensionality and focuses attention on a small number of statements, which are most likely to involve single-type faulty statements. However, for more complex programs containing multiple faulty statements or faults of different types, relying on a single failing test case for slicing is insufficient to capture the complete set of faulty statements. Therefore, we opt to use multiple failing test cases to construct a more comprehensive fault semantic context.

Thus, \appname generates a new \( M \times K^{'} \) context matrix and a new \( 1 \times K^{'} \) statement index. This context matrix integrates the set of faulty statements responsible for multiple faults by removing duplicate statements across the slices of multiple failing test cases, resulting in a comprehensive and refined representation of the faulty context.

\subsection{Statistical Context Construction}\label{rPCA}
\figurename{}~\ref{synthesis} illustrates the architecture of our model, in which we use a simplified U-Net to implement the diffusion model. The U-Net architecture includes a single downsampling layer and a single upsampling layer.

However, the fault semantic context only includes a subset of statements derived through dynamic program slicing, capturing structural and data dependencies based on specific inputs. This context is therefore limited to local information tied to particular inputs. To address this limitation, we introduce a revised PCA\cite{song2010feature}, which not only retains the statistical properties of the data but also enriches the context by incorporating statistical dependencies, complementing the structural information obtained from program slicing. Through this fusion, we ensure that the context aligns with the model’s dimensional requirements.

Algorithm ~\ref{alg_pca} describes feature selection using revised PCA. It takes the coverage matrix \( X \), the statement set \( StmSC \) from program slicing, and two key parameters: the \textit{number of largest eigenvalues \( m \)} and \textit{number of principal components \( K^{''} \)}.

It firstly computes the covariance matrix \( covX \), solves for eigenvalues and eigenvectors, and selects the top \( m \) eigenvectors (Steps 1-4).

Contribution values \( c_i \) are calculated by summing the \( m \) elements in each row of matrix \( V \), and the top indices are stored in \( iContriMax \) (Steps 6-8). 

Finally, it generates the statistical context matrix \( X_{PCA} \) of size \( M \times K^{''} \) and context index vecor \( StmPCA \) by selecting columns from \( X \) corresponding to \(  iContriMax\), and returns \( X_{PCA} \) and \( StmPCA \) (Steps 9-14).

\begin{algorithm}[!h]
    \caption{dimensionality reduction using revised PCA}
    \label{alg_pca}
    
    \KwIn{\\coverage matrix with the size of $M \times N$: $X$,\\
          number of largest eigenvalues: $m$,\\
          number of principal components: $K^{p}$}
    
    \KwOut{\\statistical context matrix with size of $M \times K^{''}$: $X_{PCA}$\\
    statistical context index with size of $1 \times K{''}$: StmPCA}
    
    $covX = $ covariance matrix of original samples\;
    $eigenVec = $ eigenvectors of $covX$\;
    $eigenVal = $ eigenvalues of $covX$\;
    $V = $ select the $eigenVec$ corresponding to the first $m$ largest $eigenVal$\;
    
    \For{$i = 1$ \KwTo $N$}{
        Calculate contribution value: $c_i = \sum_{p=1}^{m} |V_{pi}|$\;
    }
    
    $iContriMax = $ argmax($c$)\;
    Initialize $X_{PCA}$ and $StmPCA$ as None, None\;
    \For{$i = 1$ \KwTo $K^{''}$}{
        Add the iContriMax[i]-th column of $X$ to $X_{PCA}$\;  
        Add $iContriMax[i]$ to $StmPCA$\;

    }
    
    \Return{$X_{PCA}$, $StmPCA$}\;
\end{algorithm}

\begin{algorithm}[!h]
    \caption{Context Fusion Using Fault Semantic and Statistical Contexts}
    \label{alg_fusion}
    \KwIn{\\coverage matrix with the size of $M \times N$: $X$,\\
          statements index selected by program slicing with the size of $1 \times K^{'}$: $StmSC$,\\
          statistical context index with size of $1 \times K^{''}$: $StmPCA$,\\
          fusion ratio: $\alpha$}
    \KwOut{\\reduced coverage matrix with size of $M \times K$: $X_{fusion}$}

    Set fusion size as $K^{f}$ = $\alpha \times K^{'}$;\\
    Set fusion context statements index $StmFusion = StmSC \cap StmPCA[:K^{f}]$;\\  
    
    \For{$i = 1$ \KwTo $K^{''}$}{
        \If{$StmFusion$ matches the dimensional requirements of \appname or DLFL}{
            \textbf{break};
        }
    
        \If{$StmPCA[i] \notin StmFusion$}{
            \If{$StmPCA[i] \in StmSC$}{
                add $StmPCA[i]$ to $StmFusion$;
            }
        }
    }

    Initialize reduced coverage matrix $X_{fusion}$ as None;\\
    
    \For{$i = 1$ \KwTo len(StmFusion)}{
        add $StmFusion[i]-th$ column of $X$ to $X_{fusion}$\;
    }
    \Return{$X_{fusion}$}\;
\end{algorithm}

\subsection{Context-aware Diffusion Model}
Algorithm ~\ref{alg_fusion} describes the fusion process, which integrates semantic and statistical contexts to refine the coverage matrix. The inputs include the coverage matrix \( X \), the statement index \( StmSC \) from program slicing, the statistical context index \( StmPCA \), and a fusion ratio \( \alpha \).

The algorithm first calculates the fusion size \( K^f \) as \( \alpha \times K' \) and initializes \( StmFusion \) as the intersection of \( StmSC \) and the top \( K^f \) elements of \( StmPCA \) (Steps 1-2).

Next, it iterates through \( StmPCA \) to expand \( StmFusion \) as needed. If an element in \( StmPCA \) is also in \( StmSC \) but not yet in \( StmFusion \), it is added to \( StmFusion \) (Steps 3-7). This ensures that essential statements from both contexts are included.

Finally, \( X_{fusion} \) is constructed by selecting columns from \( X \) based on \( StmFusion \), and \( X_{fusion} \) is returned (Steps 8-12).

In fault localization, high-quality data augmentation must reflect the specific program contexts most relevant to causing failures. Randomly generated failing data can introduce noise, which may hinder the model's ability to effectively localize faults. To ensure that the generated data remains aligned with the original failure-inducing conditions, we explore two possible guidance strategies: classifier-based guidance and classifier-free guidance.

\subsubsection{\textbf{Classifier-Based Strategy}}

To guide sample generation in the reverse diffusion process, we leverage gradients of the target data distribution. \cite{dhariwal2021diffusion} adding a classifier gradient to the noise term can direct sample generation toward specific target classes. This approach modifies the noise prediction equation as follows: \begin{equation} \label{eq8} \hat{\epsilon} = \epsilon_{\theta}(\mathbf{x}_t) - \gamma \cdot \sqrt{1 - \bar{\alpha}t} \nabla{\mathbf{x}t} \log p_{\phi}(y | \mathbf{x}_t) \end{equation} \noindent where $\gamma$ controls the level of guidance, and $p_{\phi}(\cdot)$ represents the classifier’s probability function.

\subsubsection{\textbf{Classifier-Free Strategy}}

In fault localization, relying on a pre-trained classifier introduces extra computational cost, particularly in complex and high-dimensional program data.

To mitigate this, we adopt a classifier-free approach for guidance\cite{ho2022classifier}, which retains effective sample generation without depending on an external classifier. This alternative strategy redefines the noise prediction in the reverse process as: \begin{equation} \label{eq10} \hat{\epsilon} = (1 + \gamma) \cdot \epsilon_{\theta}(\mathbf{x}t, t, \mathbf{c}) - \gamma \cdot \epsilon{\theta}(\mathbf{x}_t, t) \end{equation} \noindent where $\gamma$ again serves as the guidance scale, offering control over the generation direction without requiring pre-trained classifier involvement.

\subsubsection{\textbf{Sampling with DPM-Solver}}

To further improve the efficiency of the diffusion process in our fault localization model, we employ the DPM-Solver sampler~\cite{lu2022dpm}. DPM-Solver is a fast, high-order solver specifically designed for diffusion ODEs, with guaranteed convergence order. It is suitable for both discrete-time and continuous-time diffusion models without requiring any further training. Experimental results demonstrate that DPM-Solver can generate high-quality samples with only 10 to 20 function evaluations across various datasets in Computer Vision. 

In our model, we set the sampling steps to \textbf{25}. DPM-Solver significantly reduces sampling time while maintaining high sample quality.  It solves the probability flow ordinary differential equation (ODE) that governs the diffusion process, approximating the reverse process as follows:

\begin{equation}
    d\mathbf{x}(t) = \epsilon_{\theta}(t) \left( \frac{\mathbf{x}(t)}{\sqrt{\sigma^2 + 1}} \right) d\sigma(t)
\end{equation}

where \(\sigma_t\) is the noise schedule parameterized by \(\sqrt{1 - \alpha_t} / \sqrt{\alpha_t}\), and \(\mathbf{x}(t)\) represents the latent state at time \(t\). This method allows our fault localization model to efficiently generate failure-related samples with minimal computational overhead, making it scalable for large program datasets with high-dimensional inputs, especially when using classifier-free guidance.

\subsection{Model Training}
After merging the fault semantic context with the statistical context, we obtain a \( M \times K \) context matrix derived from the \( M \times N \) raw data. This context matrix captures the information related to the statements that lead to program failures, combining insights from both the program's structure and statistical analysis. \appname uses this context matrix as input to the diffusion model, which generates synthesized failing test cases. The diffusion model learns the characteristics of both failing and passing test cases through its forward and reverse processes, ensuring that the newly synthesized samples reflect the key patterns present in the raw data.

In the forward process, \appname gradually adds Gaussian noise to the original context matrix over multiple time steps, following a classifier-free guidance strategy. In the reverse process, \appname predicts the noise at each step using the trained model and gradually denoises the noisy data to recover the synthesized samples. The reverse process is trained using a mean squared error (MSE) loss function to minimize the difference between the added noise and the predicted noise.

As shown in \figurename~\ref{synthesis}, the diffusion model, once trained, generates new failing test cases by sampling from noise using DPM-Solver, while also incorporating label information to ensure the synthesized samples reflect the failure characteristics of the original test cases. The newly generated failure samples are then combined with the passing data to form a balanced dataset, which is used to enhance the performance of fault localization methods such as SFL and DLFL.

\subsection{An Illustrative Example}
\begin{figure*}[htbp]
    \centerline{\includegraphics[width=0.9\textwidth]{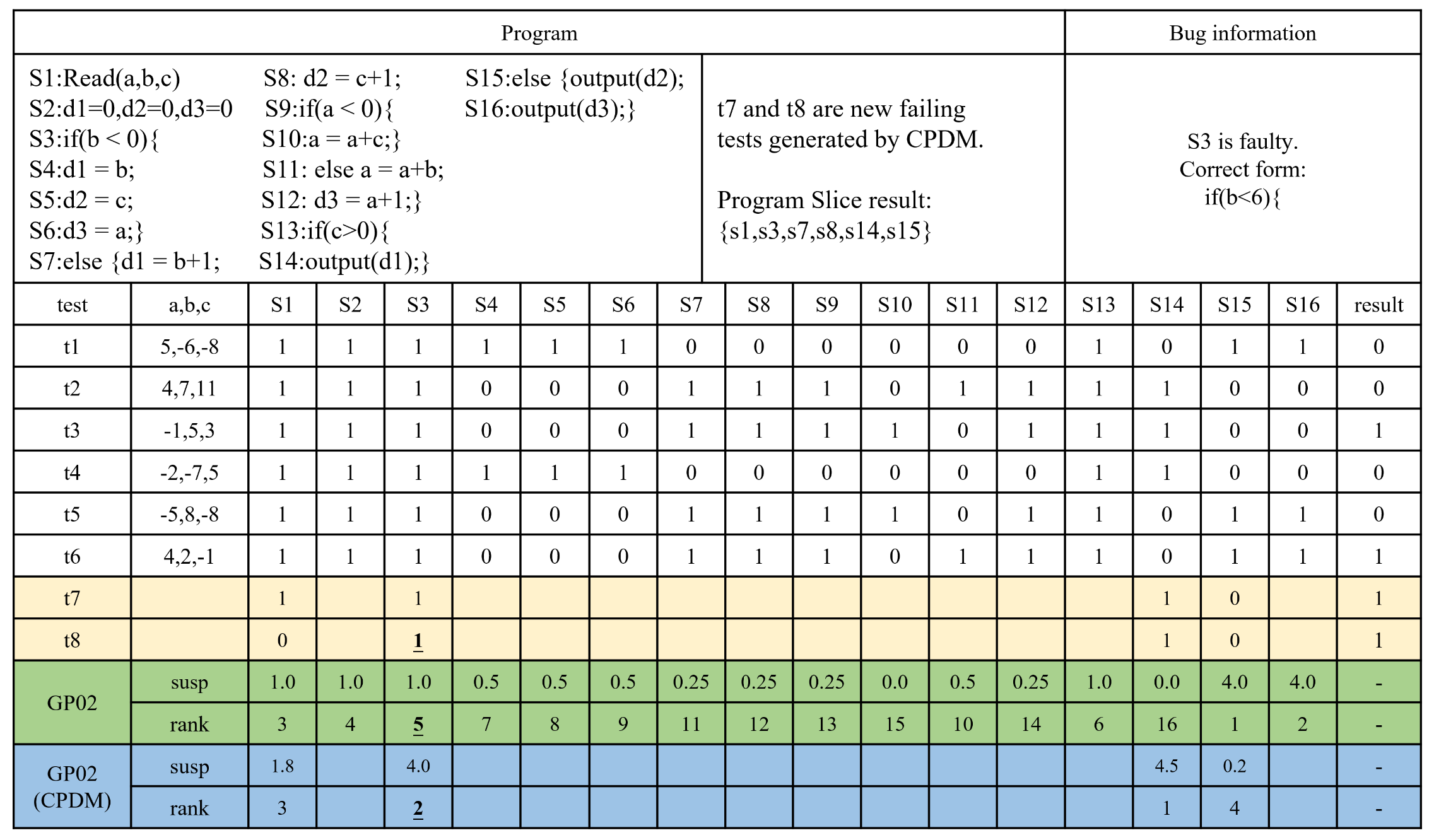}}
    \caption{An example illustrating \appname.}
    \label{example}
\end{figure*}

To illustrate the workflow of \appname, we provide an example in \figurename{} \ref{example}. Here, program \textit{P} contains 16 statements with a fault at line 3, where the value 0 is mistakenly set instead of 6. The SFL method GP02 \cite{xie2013theoretical} is applied to locate the faulty statement.

Each cell below a statement indicates its execution by a test case (0 if not executed, 1 if executed). The 'result' column in \figurename{} \ref{architecture} shows test outcomes (1 for failure, 0 for success). The original test suite is imbalanced, with four passing test cases ($t_1$, $t_2$, $t_4$, $t_5$) and two failing cases ($t_3$, $t_6$).

To balance this dataset, \appname generates two additional failing test cases. Program slicing is applied to extract fault semantic context from $t_3$ and $t_6$. Using Eq.~(\ref{slice}), we define ($S_{14}$, d1, $t_3$) and ($S_{15}$, d2, $t_6$) as slicing criteria, given the incorrect output of variable d1 at $S_{14}$ during $t_3$. As shown in \figurename{} \ref{example}, the $StmSC$ matrix for $t_3$ includes {\small$\{ S_{1}, S_{3}, S_{7}, S_{14} \}$} and for $t_6$ includes {\small$\{ S_{1}, S_{3}, S_{8}, S_{15} \}$}. So the fault semantic context contains {\small$\{ S_{1}, S_{3}, S_{7}, S_{8},S_{14}, S_{15} \}$}. Revised PCA yields the $StmPCA$ matrix {\small$\{ S_{14}, S_{15}, S_{10}, S_{11}, S_{6}, S_{1}, S{2}, S{3}, S{13}, S{4}, S{5}, S{16}, S{8}, S{7}, S{9}, S{12} \}$}. With a fusion ratio $\alpha$ set to 1, the fusion size $K^f$ is equal to $StmSC$. Fusing the fault semantic context and statistical context forms the context {\small$\{ S_{1}, S_{3}, S_{14}, S_{15} \}$}.

Using this enriched context, \appname employs a Diffusion Model to generate synthetic failing test cases ($t_7$ and $t_8$), highlighted in yellow in \figurename{} \ref{example}. These new cases expand the context matrix, allowing GP02 to re-evaluate statement suspiciousness with updated data.

The final rows of \figurename{} \ref{example} compare FL results from GP02 with and without \appname. Without \appname, GP02 ranks the statements (highlighapted in green) as {\small $\{ S_{15}, S_{16}, S_{1}, S_{2}, S_{3}, S_{13}, S_{4}, S_{5}, S_{6}, S_{11}, S_{7}, S_{8}, S_{9}, S_{12}, S_{10}, S_{14} \}$}. After applying \appname, the ranking shifts to {\small$\{ S_{14}, S_{3}, S_{1}, S_{15} \}$}.  Notably, the faulty statement $S_{3}$ moves from 5th to 2th, demonstrating \appname’s effectiveness in mitigating class imbalance and enhancing fault localization accuracy.

\section{Experiments}\label{experiments}

To assess the effectiveness of our proposed approach, we carried out experiments on 262 versions of five representative benchmark programs, all of which contain real-world faults. The selected programs—Chart, Math, Lang, Time, and Mockito—were drawn from the Defects4J\footnote{\url{https://github.com/rjust/defects4j}} dataset\cite{just2014defects4j}. Due to the substantial size of these programs, manually collecting input data would be highly time-consuming. Consequently, we leveraged the coverage matrix provided by Pearson et al.\cite{pearson2016evaluating} to optimize and expedite the experimental process.

\tablename~\ref{table1} provides an overview of the five subject programs. For each program, it includes a brief functional description (in the 'Description' column), the number of faulty versions available (in the 'Versions' column), the program size measured in thousand lines of code (in the 'LoC(K)' column), and the number of test cases (in the 'Test' column).

\begin{table}[htbp]
  \centering
  \caption{Subject programs}
    \begin{tabular}{ccccc}
    \toprule
    Programs & Description & Versions & Loc(K) & Tests \\
    \midrule
    Chart & Java chart library & 26    & 96    & 2205 \\
    Lang  & Apache commons-lang & 65    & 22    & 2245 \\
    Math  & Apache commons-math & 106   & 85    & 3602 \\
    Time  & Standard date and time library & 27    & 28    & 4130 \\
    Mockito & Mocking framework for Java & 38    & 67    & 1075 \\
    \midrule
    Total & -     & 262   & 298   & 13257 \\
    \bottomrule
    \end{tabular}
  \label{table1}
\end{table}

\subsection{Experiment Settings}
The experiments were conducted on a Linux server equipped with 40 cores of a 2.4GHz CPU and 252GB of RAM. The operating system used was Ubuntu 20.04.

\tablename~\ref{setParam} provides the main parameters used in our experiments. Notably, we applied this same set of hyper-parameters across 262 faulty versions of the programs, effectively treating each version as a unique dataset. This set demonstrates the robustness and adaptability of our approach and parameter configuration, as it performs consistently well across diverse fault scenarios.

\begin{table}[htbp]
  \centering
  \caption{Main Parameters of \appname}
    \begin{tabular}{ccc}
    \toprule
    Main Parameters & Description & Value \\
    \midrule
    steps & Number of diffusion steps & 1000 \\
    lr  & Learning rate & 0.0003 \\
    op  & Optimizer & AdamW \\
    $\beta_1$  & Initial value of beta in the diffusion process & 0.0001 \\
    $\beta_T$ & Final value of beta in the diffusion process & 0.02 \\
    $\alpha$ & Fusion ratio in context fusion & 1 \\
    \bottomrule
    \end{tabular}
  \label{setParam}
\end{table}

\subsection{Evaluation Metrics}
We employ four widely recognized metrics in FL to evaluate the performance of our approach:

\begin{itemize}
    \item \textbf{Number of Top-K}: It quantifies the number of faulty versions where at least one fault is ranked within the top $K$ positions by a fault localization (FL) method. Following the prior work\cite{kochhar2016practitioners, li2021fault}, we assign K with the value of 1, 3 and 5 for our evaluation.
    \item \textbf{Mean Average Rank (MAR)\cite{li2021fault}}: For each faulty version, we calculate the average rank of all faulty statements in the ranking list. A lower value of MAR indicates better FL effectiveness.
    \item \textbf{Mean First Rank (MFR)\cite{li2021fault}}: MFR determines the rank of the first located faulty statement for each version and computes the mean rank across all versions.
   \item \textbf{Relative Improvement (RImp)\cite{zhang2019cnn}}: This metric evaluates the efficiency of fault localization methods by comparing the number of statements that must be examined. RImp specifically reflects the proportion of statements examined when using our method in comparison to others, where a lower RImp value indicates superior performance. 

\end{itemize}

These metrics offer a thorough evaluation of our approach's fault localization accuracy, allowing for performance comparisons with other fault localization methods.

\subsection{Research Questions and Results}
We evaluate the effectiveness of our approach through the following four research questions.
\\

\noindent\textbf{\textit{RQ1. How effective is \appname in localizing real faults compared with original state-of-the-art SFL methods?}}

We assessed the performance of three statement-level SFL methods (Dstar\cite{wong2012software}, Ochiai\cite{abreu2006evaluation}, and Barinel\cite{abreu2009spectrum}) under two different conditions: the original SFL method and our approach \appname version. These original methods typically process raw data without any additional context. The results for Top-1, Top-3, Top-5, MFR, and MAR metrics are summarized in \tablename~\ref{table_RQ1_1}, offering a comparison between the original methods and our approach \appname.

\begin{table}[htbp]
  \centering
  \caption{The results of TOP-1, TOP-3, TOP-5, MFR and MAR by comparison of original SFL method and \textbf{\appname}.}
    \resizebox{\textwidth}{!}{
    \begin{tabular}{lllllllllllllllllllll}
    \toprule
    \multirow{2}[4]{*}{Program} & \multirow{2}[4]{*}{Scenario} & \multicolumn{3}{l}{Top-1} &       & \multicolumn{3}{l}{Top-3} &       & \multicolumn{3}{l}{Top-5} &       & \multicolumn{3}{l}{MFR} &       & \multicolumn{3}{l}{MAR} \\
\cmidrule{3-5}\cmidrule{7-9}\cmidrule{11-13}\cmidrule{15-17}\cmidrule{19-21}          &       & Dstar & Ochiai & Barinel &       & Dstar & Ochiai & Barinel &       & Dstar & Ochiai & Barinel &       & Dstar & Ochiai & Barinel &       & Dstar & Ochiai & Barinel \\
    \midrule
    \multirow{2}[2]{*}{Chart} & Origin & \textbf{3} & \textbf{3} & \textbf{2} &       & \textbf{7} & 7     & 6     &       & \textbf{10} & \textbf{10} & \textbf{10} &       & 305.74  & 189.26  & 138.13  &       & 322.49  & 206.93  & 152.35  \\
          & PCD-DAug & \textbf{3} & \textbf{3} & 1     &       & \textbf{7} & \textbf{8} & \textbf{7} &       & 9     & 9     & \textbf{10} &       & \textbf{116.25 } & \textbf{116.46 } & \textbf{66.00 } &       & \textbf{125.19 } & \textbf{125.31 } & \textbf{72.30 } \\
    \midrule
    \multirow{2}[2]{*}{Lang} & Origin & 5     & 5     & 5     &       & 17    & 17    & 17    &       & 24    & 24    & 24    &       & 29.14  & 30.45  & 33.17  &       & 81.61  & 60.41  & 61.48  \\
          & PCD-DAug & \textbf{12} & \textbf{12} & \textbf{10} &       & \textbf{19} & \textbf{19} & \textbf{21} &       & \textbf{27} & \textbf{28} & \textbf{28} &       & \textbf{19.00 } & \textbf{18.78 } & \textbf{23.00 } &       & \textbf{27.67 } & \textbf{27.67 } & \textbf{29.32 } \\
    \midrule
    \multirow{2}[2]{*}{Math} & Origin & 17    & \textbf{17} & 17    &       & \textbf{32} & \textbf{32} & \textbf{31} &       & 38    & \textbf{39} & 39    &       & \textbf{59.84 } & \textbf{61.35 } & 68.24  &       & 315.93  & 209.81  & 212.46  \\
          & PCD-DAug & \textbf{18} & 15    & \textbf{18} &       & \textbf{32} & \textbf{32} & 30    &       & \textbf{40} & \textbf{39} & \textbf{40} &       & 65.48  & 66.02  & \textbf{64.04 } &       & \textbf{98.97 } & \textbf{99.74 } & \textbf{93.07 } \\
    \midrule
    \multirow{2}[2]{*}{Time} & Origin & 2     & 2     & \textbf{2} &       & \textbf{10} & \textbf{10} & \textbf{9} &       & \textbf{10} & \textbf{10} & \textbf{10} &       & 398.63  & 598.96  & 600.78  &       & 675.20  & 874.50  & 875.55  \\
          & PCD-DAug & \textbf{3} & \textbf{3} & 1     &       & 8     & 7     & 8     &       & \textbf{10} & \textbf{10} & \textbf{10} &       & \textbf{125.89 } & \textbf{126.74 } & \textbf{135.78 } &       & \textbf{137.04 } & \textbf{137.53 } & \textbf{143.47 } \\
    \midrule
    \multirow{2}[2]{*}{Mokito} & Origin & \textbf{5} & \textbf{6} & 4     &       & \textbf{11} & \textbf{11} & \textbf{8} &       & 12    & 12    & 10    &       & 276.86  & 220.06  & 230.08  &       & 461.74  & 417.09  & 420.42  \\
          & PCD-DAug & \textbf{5} & 5     & \textbf{6} &       & 8     & 9     & \textbf{8} &       & \textbf{12} & \textbf{12} & \textbf{10} &       & \textbf{94.39 } & \textbf{92.29 } & \textbf{90.71 } &       & \textbf{132.01 } & \textbf{129.91 } & \textbf{125.49 } \\
    \midrule
    \multirow{2}[2]{*}{Total} & Origin & 32    & 33    & 30    &       & \textbf{77} & \textbf{77} & 71    &       & 94    & 95    & 93    &       & 140.07  & 143.69  & 144.24  &       & 315.42  & 270.63  & 267.69  \\
          & PCD-DAug & \textbf{41} & \textbf{38} & \textbf{36} &       & 74    & 75    & \textbf{74} &       & \textbf{98} & \textbf{98} & \textbf{98} &       & \textbf{69.05 } & \textbf{69.01 } & \textbf{65.31 } &       & \textbf{92.35 } & \textbf{92.41 } & \textbf{85.19 } \\
    \bottomrule
    \end{tabular} }
  \label{table_RQ1_1}%
\end{table}%

\textit{\textbf{Top-K. }}As shown in \tablename~\ref{table_RQ1_1}, \appname demonstrates a clear advantage in Top-K metrics across different program datasets, with particularly strong performance on the Lang dataset. In the Top-1, Top-3, and Top-5 rankings, \appname identified 11, 19, and 27 faults on average using three SFL methods (Dstar, Ochiai, and Barinel), compared to 5, 17, and 24 faults identified by the original methods. This represents performance improvements of 120\%, 11.76\%, and 12.5\%, respectively, indicating that the hyper-parameters used in \appname were especially effective for the Lang dataset.

In other datasets, \appname generally outperforms the original methods as well. For instance, in the Math dataset, \appname shows a slight average improvement across all Top-K metrics. However, on the Mokito dataset, \appname's performance in Top-3 metrics slightly lags behind the original methods. Specifically, the average number of faults identified in the Top-3 rankings decreased by 16.67\%, respectively. This may be due to the hyper-parameters being optimized for the Lang dataset, without further adjustment for the specific characteristics of other datasets, which could have led to a slight performance trade-off on certain datasets.

Overall, \appname exhibits consistent performance gains across all SFL methods in Top-1 and Top-5 metrics. For example, using the Barinel method, \appname correctly identified 36, 74, and 98 faults in the Top-1, Top-3, and Top-5 rankings, compared to 30, 71, and 93 faults identified by the original Barinel method. This corresponds to improvements of 20.00\%, 4.23\%, and 5.38\%, respectively. These results demonstrate that \appname not only performs well on individual datasets but also shows superior average performance across all datasets, validating its effectiveness in improving fault localization accuracy.

Furthermore, the use of unified hyper-parameters brings important advantages to the application of \appname. First, it simplifies the model deployment process by eliminating the need to tune hyper-parameters for each individual dataset, thereby enhancing the model's generality. Second, using a unified hyper-parameter setting helps reduce the risk of overfitting, contributing to greater stability and consistency in model performance. Finally, this unified configuration enhances the reproducibility of experimental results and allows for better performance comparison across datasets.

\begin{figure}[htbp]    
\centerline{\includegraphics[scale=0.6]{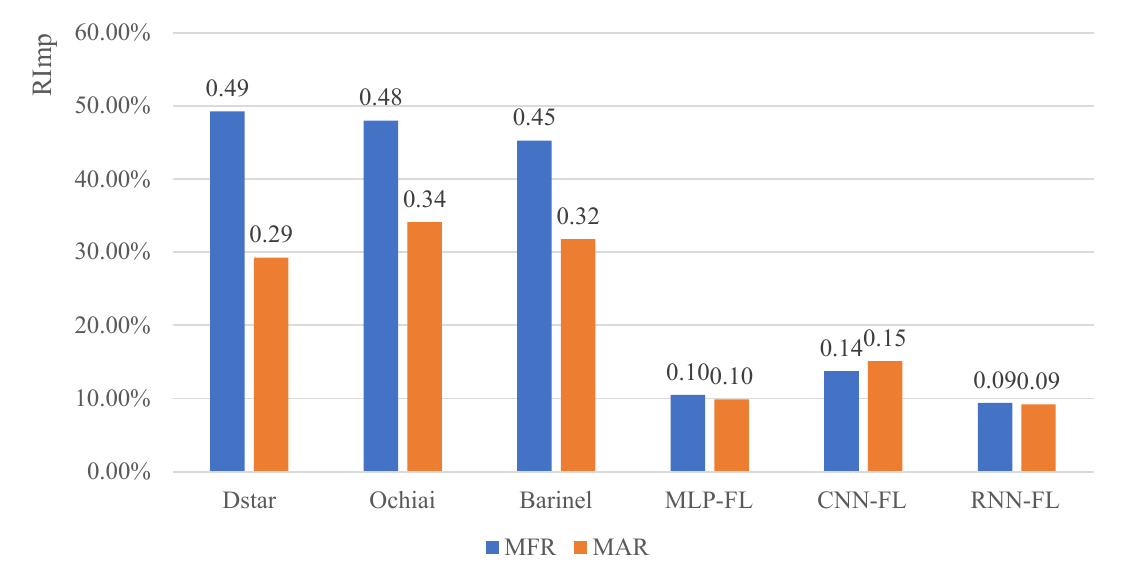}}
    \caption{The RImp of MFR and MAR for \appname over six original FL methods.}
    \label{fig_RQ12_RImp}
\end{figure}

\textit{\textbf{RImp.}} In \figurename~\ref{fig_RQ12_RImp}, the RImp values across the evaluated SFL methods (Dstar, Ochiai, and Barinel) consistently remain below 100\%, demonstrating that \appname outperforms these traditional techniques in fault localization efficiency. For example, as depicted in \figurename~\ref{fig_RQ12_RImp}, \appname significantly reduces the MFR metric across the SFL methods. When using \appname, the percentage of statements that need to be inspected to find first faulty statement ranges from 45.28\% with Barinel to 49.29\% with Dstar. This implies that \appname can reduce the number of statements required for examination by 50.71\% ($100\% - 49.29\% = 50.71\%$) for Dstar and 54.72\% ($100\% - 45.28\% = 54.72\%$) for Barinel. Thus, our approach \appname can lead to substantial reductions in the effort required for fault localization.



\begin{tcolorbox}[boxrule=0pt, frame empty, sharp corners, left=0mm, right=0mm]
 \emph{\textbf{Summary for RQ1}: In RQ1, we evaluated the performance of \appname against three traditional SFL methods. The findings show that \appname achieves better results compared to the original methods, indicating that \appname provides a more effective approach for fault localization.
}
\end{tcolorbox}

\noindent\textbf{\textit{RQ2. How effective is \appname in localizing real faults compared with the state-of-the-art DLFL methods?}}

\begin{table*}[htbp]
  \centering
  \caption{The results of TOP-1, TOP-3, TOP-5, MFR and MAR by comparison of original DLFL method and \textbf{\appname}.}
  \resizebox{\textwidth}{!}{
    \begin{tabular}{lllllllllllllllllllll}
    \toprule
    \multirow{2}[4]{*}{Program} & \multirow{2}[4]{*}{Scenario} & \multicolumn{3}{l}{Top-1} &       & \multicolumn{3}{l}{Top-3} &       & \multicolumn{3}{l}{Top-5} &       & \multicolumn{3}{l}{MFR} &       & \multicolumn{3}{l}{MAR} \\
\cmidrule{3-5}\cmidrule{7-9}\cmidrule{11-13}\cmidrule{15-17}\cmidrule{19-21}          &       & MLP-FL & CNN-FL & RNN-FL &       & MLP-FL & CNN-FL & RNN-FL &       & MLP-FL & CNN-FL & RNN-FL &       & MLP-FL & CNN-FL & RNN-FL &       & MLP-FL & CNN-FL & RNN-FL \\
    \midrule
    \multirow{2}[2]{*}{Chart} & Origin & 1     & 0     & 2     &       & 3     & 0     & 3     &       & 4     & 0     & 3     &       & 609.22  & 560.22  & 848.78  &       & 807.25  & 649.92  & 967.27  \\
          & PCD-DAug & \textbf{5} & \textbf{1} & \textbf{3} &       & \textbf{8} & \textbf{1} & \textbf{6} &       & \textbf{10} & \textbf{3} & \textbf{9} &       & \textbf{63.21 } & \textbf{158.79 } & \textbf{68.58 } &       & \textbf{68.07 } & \textbf{178.46 } & \textbf{86.08 } \\
    \midrule
    \multirow{2}[2]{*}{Lang} & Origin & 3     & 0     & 0     &       & 8     & 1     & 2     &       & 13    & 2     & 4     &       & 196.43  & 390.43  & 331.34  &       & 289.21  & 442.95  & 442.47  \\
          & PCD-DAug & \textbf{13} & \textbf{7} & \textbf{8} &       & \textbf{20} & \textbf{11} & \textbf{15} &       & \textbf{28} & \textbf{19} & \textbf{25} &       & \textbf{18.92 } & \textbf{23.00 } & \textbf{26.51 } &       & \textbf{30.55 } & \textbf{30.27 } & \textbf{33.84 } \\
    \midrule
    \multirow{2}[2]{*}{Math} & Origin & 1     & 2     & 0     &       & 10    & 2     & 3     &       & 12    & 2     & 6     &       & 526.78  & 1104.67  & 950.20  &       & 978.65  & 1190.07  & 1316.76  \\
          & PCD-DAug & \textbf{19} & \textbf{5} & \textbf{6} &       & \textbf{31} & \textbf{11} & \textbf{14} &       & \textbf{36} & \textbf{16} & \textbf{22} &       & \textbf{64.89 } & \textbf{118.63 } & \textbf{70.63 } &       & \textbf{102.27 } & \textbf{152.97 } & \textbf{103.28 } \\
    \midrule
    \multirow{2}[2]{*}{Time} & Origin & 0     & \textbf{0} & 0     &       & 0     & \textbf{0} & 0     &       & 0     & \textbf{0} & 0     &       & 2228.56  & 2603.96  & 1627.07  &       & 2756.34  & 2809.59  & 2211.87  \\
          & PCD-DAug & \textbf{1} & \textbf{0} & \textbf{1} &       & \textbf{2} & \textbf{0} & \textbf{2} &       & \textbf{5} & \textbf{0} & \textbf{3} &       & \textbf{129.15 } & \textbf{460.63 } & \textbf{173.14 } &       & \textbf{162.22 } & \textbf{508.83 } & \textbf{198.90 } \\
    \midrule
    \multirow{2}[2]{*}{Mokito} & Origin & 0     & \textbf{0} & 0     &       & 0     & 0     & 0     &       & 0     & 1     & 0     &       & 527.44  & 811.97  & 578.86  &       & 808.75  & 1005.73  & 890.16  \\
          & PCD-DAug & \textbf{4} & \textbf{0} & \textbf{3} &       & \textbf{6} & \textbf{2} & \textbf{4} &       & \textbf{8} & \textbf{4} & \textbf{5} &       & \textbf{107.11 } & \textbf{136.26 } & \textbf{109.58 } &       & \textbf{149.02 } & \textbf{182.05 } & \textbf{149.83 } \\
    \midrule
    \multirow{2}[2]{*}{Total} & Origin & 5     & 2     & 2     &       & 21    & 3     & 8     &       & 29    & 5     & 13    &       & 629.49  & 991.81  & 803.70  &       & 951.90  & 1097.09  & 1098.64  \\
          & PCD-DAug & \textbf{42} & \textbf{13} & \textbf{21} &       & \textbf{67} & \textbf{25} & \textbf{41} &       & \textbf{87} & \textbf{42} & \textbf{64} &       & \textbf{66.08 } & \textbf{136.52 } & \textbf{75.81 } &       & \textbf{94.24 } & \textbf{165.85 } & \textbf{101.07 } \\
    \bottomrule
    \end{tabular} }
  \label{table_RQ2_1}%
\end{table*}

In addition to comparing \appname with the original SFL methods, we also evaluated its performance against three representative DLFL approaches: MLP-FL\cite{zheng2016fault}, CNN-FL\cite{zhang2019cnn}, and RNN-FL\cite{zhang2021study}. As shown in \tablename~\ref{table_RQ2_1}, \appname consistently outperforms these methods across all Top-K metrics.

\textbf{\textit{Top-K.}} For example, in comparison with MLP-FL, \appname located 42, 67, and 87 faults in the Top-1, Top-3, and Top-5 metrics, respectively, while MLP-FL only identified 5, 21, and 29 faults in these categories. This represents substantial improvements of 740.00\%, 219.05\%, and 200.00\% in Top-1, Top-3, and Top-5 metrics, respectively, for \appname over the original MLP-FL method.

\textbf{\textit{RImp.}} Furthermore, \appname achieves lower mean first rank (MFR) and mean average rank (MAR) values compared to all baseline DLFL methods, indicating a more efficient fault localization process. As illustrated in \figurename~\ref{fig_RQ12_RImp}, the RImp values for MFR reveal that \appname significantly reduces the number of statements requiring inspection. With \appname, the statements needing examination range from 9.43\% (for RNN-FL) to 13.76\% (for CNN-FL), corresponding to reductions of 86.24\% ($100\% - 13.76\% = 86.24\%$) to 90.57\% ($100\% - 9.43\% = 90.57\%$) in comparison to the original DLFL approaches.

Similarly, for the MAR metric, \appname reduces the number of statements to be examined to between 9.20\% (for RNN-FL) and 15.12\% (for CNN-FL), translating to reductions of 84.88\% ($100\% - 15.12\% = 84.88\%$) to 90.80\% ($100\% - 9.20\% = 90.80\%$). These results demonstrate \appname’s substantial efficiency gains, significantly minimizing the fault localization effort compared to the original DLFL methods.

\begin{tcolorbox}[boxrule=0pt, frame empty, sharp corners, left=0mm, right=0mm] \emph{\textbf{Summary for RQ2}: In RQ2, we analyzed the performance of \appname against three DLFL methods. The result reveals that \appname consistently outperforms these methods across most metrics. These findings demonstrate that \appname is more effective at fault localization compared to the state-of-the-art DLFL methods.} \end{tcolorbox}

\noindent\textbf{\textit{RQ3. How effective is \appname in localizing real faults compared with the data optimization FL methods?}}

In addition to comparing \appname with the original SFL and DLFL methods, we evaluated its performance against two widely-used data optimization techniques: undersampling\cite{wang2020ietcr} and resampling\cite{gao2013theoretical, zhang2017theoretical,zhang2021improving}. As shown in \tablename~\ref{table_RQ3_1}, \appname consistently outperforms both optimization methods across all Top-K metrics. For example, using the Barinel fault localization method, \appname located 36, 74, and 98 faults in the Top-1, Top-3, and Top-5 metrics, respectively. In comparison, the undersampling method identified only 15, 40, and 62 faults, while the resampling method located 29, 68, and 89 faults. This translates to \appname achieving improvements of 106.67\%, 85.00\%, and 58.06\% over undersampling for the Top-1, Top-3, and Top-5 metrics, respectively, and surpassing resampling by 24.14\%, 8.82\%, and 10.11\% in these same metrics. These results highlight \appname’s enhanced fault localization effectiveness over both data optimization techniques.

\begin{table*}[htbp]
  \centering
  \caption{Comparisons between \appname and two data optimization methods for TOP-1, TOP-3, TOP-5, MAR, and MFR.}
  \resizebox{\textwidth}{!}{
    \begin{tabular}{lllllll}
    \toprule
    FL    & Scenario & Top-1 & Top-3 & Top-5 & MFR   & MAR \\
    \midrule
    \multirow{3}[2]{*}{Dstar} & Undersampling & 16    & 42    & 62    & 185.08  & 346.63  \\
          & Resampling & 28    & 66    & 82    & 154.68  & 325.00  \\
          & PCD-DAug & \textbf{41} & \textbf{74} & \textbf{98} & \textbf{69.05 } & \textbf{92.35 } \\
    \midrule
    \multirow{3}[2]{*}{Ochiai} & Undersampling & 16    & 42    & 62    & 187.27  & 327.63  \\
          & Resampling & 28    & 66    & 82    & 156.41  & 279.12  \\
          & PCD-DAug & \textbf{38} & \textbf{75} & \textbf{98} & \textbf{69.01 } & \textbf{92.41 } \\
    \midrule
    \multirow{3}[2]{*}{Barinel} & Undersampling & 15    & 40    & 62    & 175.89  & 308.74  \\
          & Resampling & 29    & 68    & 89    & 140.76  & 256.50  \\
          & PCD-DAug & \textbf{36} & \textbf{74} & \textbf{98} & \textbf{65.31 } & \textbf{85.19 } \\
    \midrule
    \multirow{3}[2]{*}{MLP-FL} & Undersampling & 20    & 47    & 63    & 230.83  & 493.45  \\
          & Resampling & 33    & \textbf{67} & 84    & 244.83  & 539.48  \\
          & PCD-DAug & \textbf{42} & \textbf{67} & \textbf{87} & \textbf{66.08 } & \textbf{94.24 } \\
    \midrule
    \multirow{3}[2]{*}{CNN-FL} & Undersampling & 1     & 6     & 9     & 997.39  & 1123.55  \\
          & Resampling & 1     & 2     & 4     & 845.84  & 1002.27  \\
          & PCD-DAug & \textbf{13} & \textbf{25} & \textbf{42} & \textbf{136.52 } & \textbf{165.85 } \\
    \midrule
    \multirow{3}[2]{*}{RNN-FL} & Undersampling & 2     & 14    & 29    & 362.07  & 639.37  \\
          & Resampling & 13    & 30    & 47    & 275.63  & 464.28  \\
          & PCD-DAug & \textbf{21} & \textbf{44} & \textbf{64} & \textbf{75.81 } & \textbf{101.07 } \\
    \bottomrule
    \end{tabular} }
  \label{table_RQ3_1}%
\end{table*}%

\begin{figure}[htbp]    
\centerline{\includegraphics[scale=0.5]{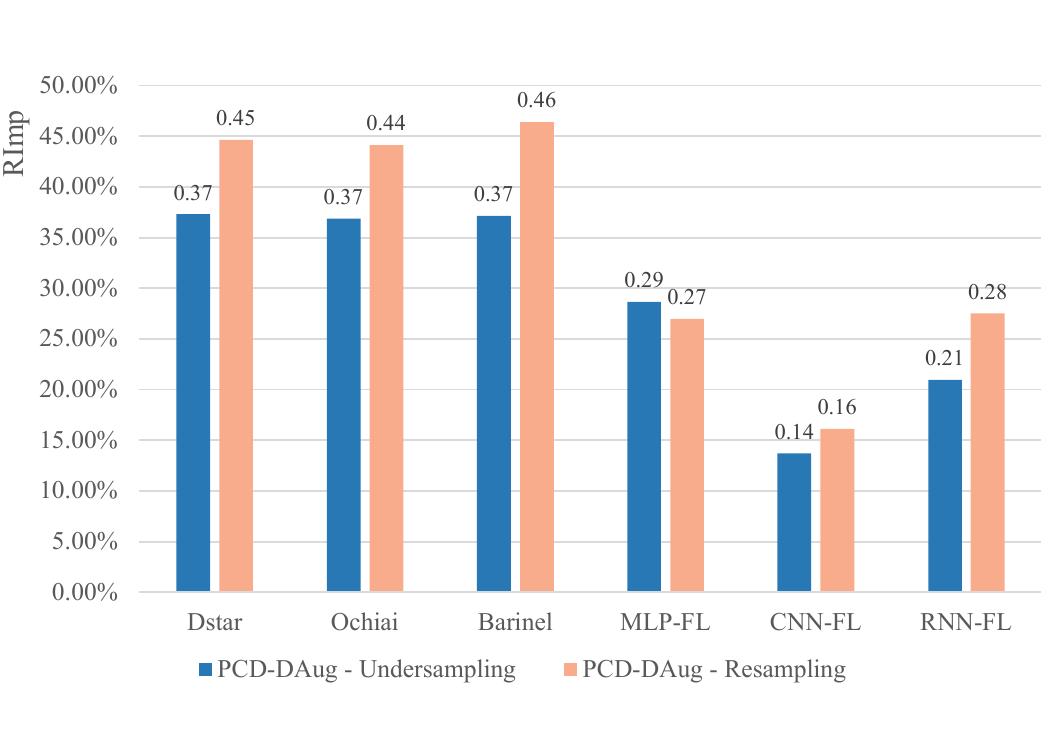}}
    \caption{The RImp of MFR by \appname over two data optimization methods.}
    \label{fig_RQ3_RImp_MFR}
\end{figure}

\begin{figure}[htbp]    
\centerline{\includegraphics[scale=0.5]{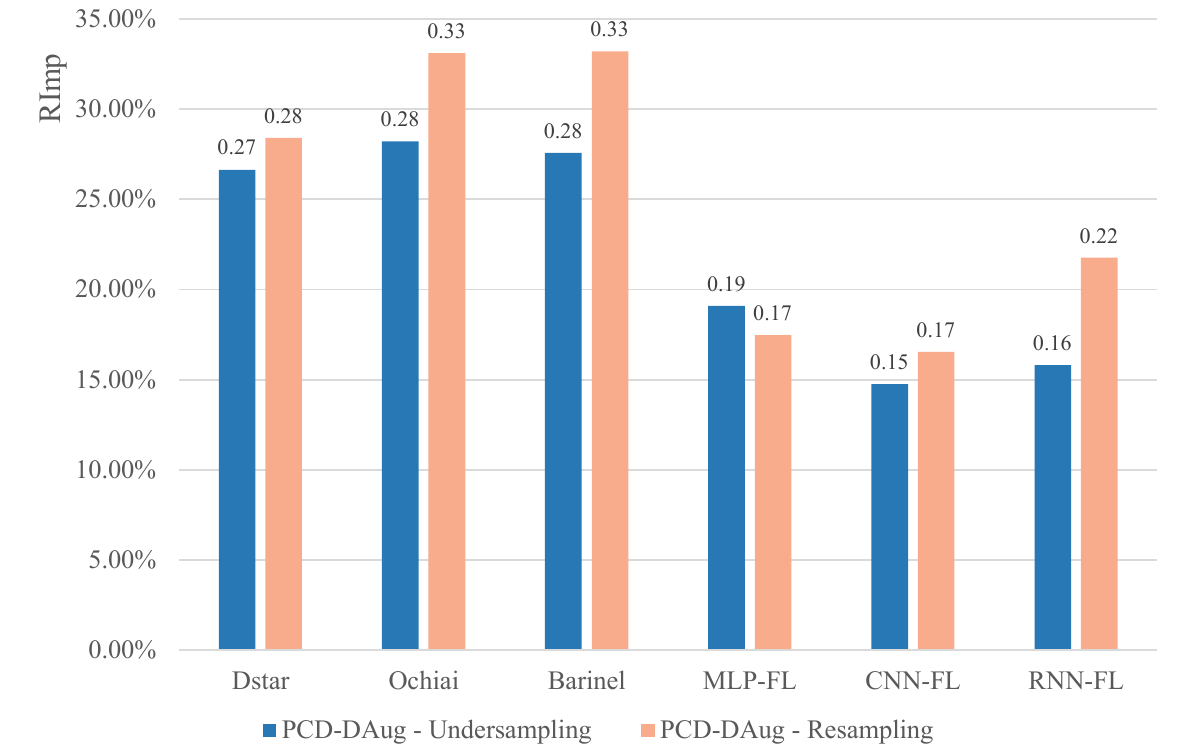}}
    \caption{The RImp of MAR by \appname over two data optimization methods.}
    \label{fig_RQ3_RImp_MAR}
\end{figure}

Moreover, in terms of the MFR and MAR metrics, \appname achieves lower values than both the undersampling and resampling methods, indicating that it ranks faulty statements higher on average and requires fewer statements to be inspected to locate faults. This efficiency in fault localization is further supported by the RImp values, as shown in \figurename~\ref{fig_RQ3_RImp_MFR} and \figurename~\ref{fig_RQ3_RImp_MAR}. All RImp values are below 100\%, indicating that \appname requires fewer statements to be examined than either undersampling or resampling.

Specifically, for the MFR metric, \appname reduces the number of statements that need to be examined to between 13.69\% (for CNN-FL) and 37.31\% (for Dstar) compared to undersampling, equating to reductions of 62.69\% ($100\% - 37.31\%$) to 86.31\% ($100\% - 13.69\%$). When compared to resampling, \appname reduces the statements to be examined to between 16.14\% (for CNN-FL) and 46.40\% (for Barinel), corresponding to reductions of 53.60\% ($100\% - 46.40\%$) to 83.86\% ($100\% - 16.14\%$).

Similarly, for the MAR metric, \appname consistently requires fewer statements to be inspected across all six fault localization methods compared to both undersampling and resampling. \appname reduces the average number of statements that need to be examined to between 14.76\% (for CNN-FL) and 28.21\% (for Ochihai) compared to undersampling, equating to reductions of 71.79\% ($100\% - 28.21\%$) to 85.24\% ($100\% - 14.76\%$). When compared to resampling, \appname reduces the statements to be examined to between 16.55\% (for CNN-FL) and 33.11\% (for Ochiai), corresponding to reductions of 76.89\% ($100\% - 33.11\%$) to 83.45\% ($100\% - 16.55\%$). This significant reduction in the number of statements inspected demonstrates \appname’s consistent advantage in minimizing the fault localization effort. These findings collectively validate the superior effectiveness and efficiency of \appname in fault localization tasks.

\begin{tcolorbox}[boxrule=0pt, frame empty, sharp corners, left=0mm, right=0mm] \emph{\textbf{Summary for RQ3}: In RQ3, we evaluated the performance of \appname against two data optimization techniques. The analysis shows that \appname surpasses both undersampling and resampling in terms of fault localization, with improvements observed across most metrics. This demonstrates that \appname is a more effective approach for fault localization compared to these optimization methods.} \end{tcolorbox}

\noindent\textbf{\textit{RQ4. How effective is \appname in localizing real faults compared with four state-of-the-art data augmentation methods?}}

In addition to comparing \appname with data optimization methods, we also evaluated its performance against four data augmentation approaches: Aeneas \cite{xie2022universal}, Lamont \cite{hu2023light}, CGAN4FL \cite{lei2023mitigating}, and PRAM \cite{hu2024deep}.

\begin{table*}[htbp]
  \centering
  \caption{Comparisons between \appname and four data augmentation methods for TOP-1, TOP-3, TOP-5, MAR, and MFR.}
  \resizebox{\textwidth}{!}{
    \small 
    \begin{tabular}{l p{2cm} p{1cm} p{1cm} p{1cm} p{2cm} p{2cm}}
    \toprule
    FL    & Scenario & Top-1 & Top-3 & Top-5 & MFR   & MAR \\
    \midrule
    Dstar & Aeneas & 12    & 31    & 39    & 489.19  & 657.84  \\
          & Lamont & 31    & 61    & 74    & 294.93  & 495.04  \\
          & CGAN4FL & 35    & 72    & 87    & 78.44  & 109.18  \\
          & PRAM  & 30    & \textbf{74} & \textbf{99} & 73.38  & 93.06  \\
          & PCD-DAug & \textbf{41} & \textbf{74} & 98    & \textbf{69.05 } & \textbf{92.35 } \\
    \midrule
    Ochiai & Aeneas & 30    & 56    & 70    & 462.96  & 647.84  \\
          & Lamont & 31    & 61    & 73    & 279.83  & 449.67  \\
          & CGAN4FL & 34    & 70    & 83    & 94.73  & 130.45  \\
          & PRAM  & 31    & \textbf{75} & \textbf{98} & 72.56  & 96.12  \\
          & PCD-DAug & \textbf{38} & \textbf{75} & \textbf{98} & \textbf{69.01 } & \textbf{92.41 } \\
    \midrule
    Barinel & Aeneas & 18    & 38    & 50    & 475.02  & 656.98  \\
          & Lamont & 30    & 60    & 74    & 265.61  & 441.02  \\
          & CGAN4FL & 28    & 55    & 66    & 89.38  & 116.50  \\
          & PRAM  & 30    & \textbf{74} & 97    & 73.41  & 96.69  \\
          & PCD-DAug & \textbf{36} & \textbf{74} & \textbf{98} & \textbf{65.31 } & \textbf{85.19 } \\
    \midrule
    MLP-FL & Aeneas & 8     & 24    & 32    & 477.75  & 655.15  \\
          & Lamont & 27    & 60    & 79    & 341.09  & 614.91  \\
          & CGAN4FL & 31    & 58    & 67    & 100.86  & 144.63  \\
          & PRAM  & 24    & 65    & 80    & 68.35  & 106.70  \\
          & PCD-DAug & \textbf{42} & \textbf{67} & \textbf{87} & \textbf{66.08 } & \textbf{94.24 } \\
    \midrule
    CNN-FL & Aeneas & 1     & 7     & 12    & 516.34  & 691.63  \\
          & Lamont & 1     & 2     & 4     & 691.87  & 905.89  \\
          & CGAN4FL & 2     & \textbf{26}    & 39    & 170.08  & 197.94  \\
          & PRAM  & 5     & 24    & 32    & 173.92  & 221.50  \\
          & PCD-DAug & \textbf{13} & 25 & \textbf{42} & \textbf{136.52 } & \textbf{165.85 } \\
    \midrule
    RNN-FL & Aeneas & 2     & 6     & 11    & 489.77  & 673.54  \\
          & Lamont & 9     & 26    & 33    & 399.58  & 665.69  \\
          & CGAN4FL & 15    & 32    & 45    & 101.60  & 142.19  \\
          & PRAM  & \textbf{22} & \textbf{45} & \textbf{65} & 95.69  & 132.16  \\
          & PCD-DAug & 21    & 44    & 64    & \textbf{75.81 } & \textbf{101.07 } \\
    \bottomrule
    \end{tabular} }
  \label{table_RQ4_1}%
\end{table*}%

As shown in \tablename~\ref{table_RQ4_1}, \appname consistently outperforms the other data augmentation methods across all cases based on Top-K metrics and the MFR and MAR rankings except for CGAN4FL in CNN-FL on Top-3 metric and PRAM in RNN-FL. Specifically, \appname achieves higher fault identification rates at Top-1, Top-3, and Top-5 levels across all fault localization (FL) methods. For instance, in the Barinel method, \appname identified 36, 74, and 98 faults at the Top-1, Top-3, and Top-5 levels, respectively, outperforming CGAN4FL (28, 55, 66) and PRAM (30, 74, 97). This translates to improvements of 28.57\%, 34.55\%, and 48.48\% over CGAN4FL, and 20.00\%, 0.00\%, and 2.06\% over PRAM, demonstrating a significant increase in fault localization accuracy.

\appname's advantage also extends to efficiency metrics. In terms of Mean First Rank (MFR) and Mean Average Rank (MAR), \appname shows consistently lower scores across all FL methods, indicating that it requires fewer statements to be examined. For example, in the Barinel method, \appname's MFR is 65.31, significantly lower than CGAN4FL's 89.38 and PRAM's 73.41, which indicates faster fault detection. In the MAR metric, \appname also performs well; for instance, in the CNN-FL method, \appname achieves a MAR of 165.85, much lower than PRAM's 221.50. This demonstrates that \appname not only improves fault localization accuracy but also reduces the effort needed for locating faults.

\begin{figure}[htbp]    
\centerline{\includegraphics[scale=0.5]{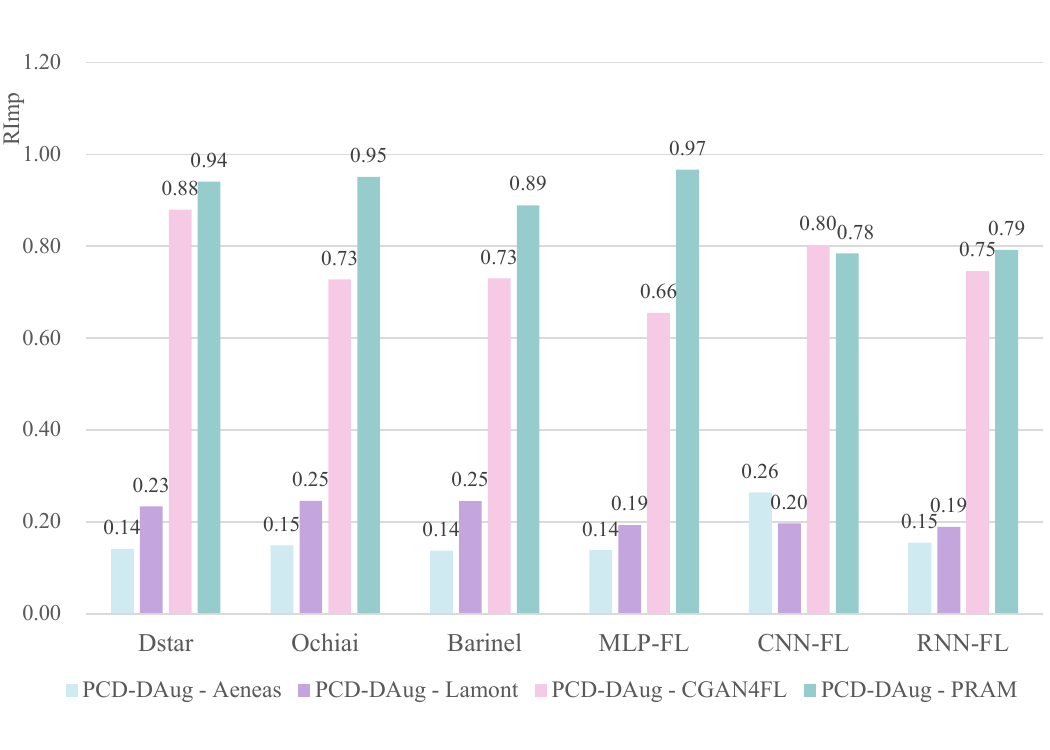}}
    \caption{The RImp of MFR by \appname over four data augmentation methods.}
    \label{fig_RQ4_RImp_MFR}
\end{figure}

\begin{figure}[htbp]    
\centerline{\includegraphics[scale=0.5]{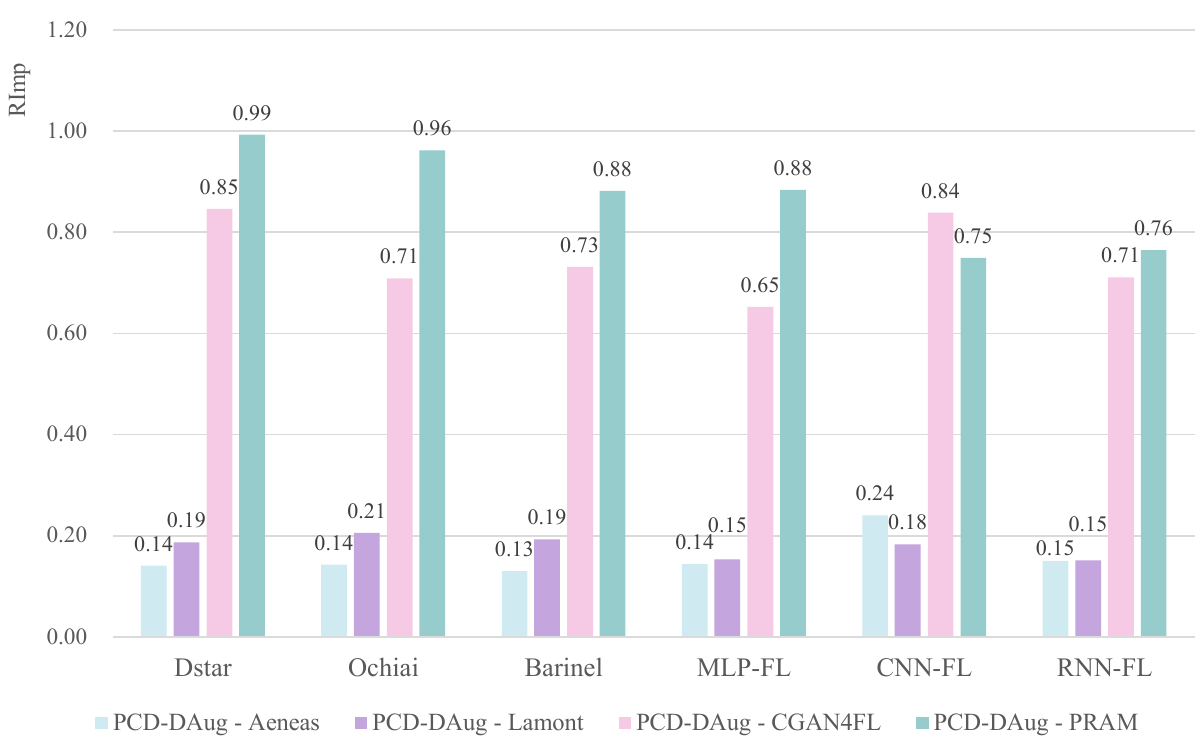}}
    \caption{The RImp of MAR by \appname over four data augmentation methods.}
    \label{fig_RQ4_RImp_MAR}
\end{figure}

\figurename~\ref{fig_RQ4_RImp_MFR} and \figurename~\ref{fig_RQ4_RImp_MAR} further illustrate \appname's relative improvement (RImp) o

.
ver Aeneas, Lamont, CGAN4FL, and PRAM in both MFR and MAR metrics across six different FL methods. In all cases, the RImp values are below 100\%, indicating that \appname requires fewer statements to be examined than other data augmentation methods. Specifically, for the MFR metric, \appname reduces the number of statements to be examined to between 78.50\% (for CNN-FL) and 95.11\% (for Ochiai) compared to PRAM, representing reductions of 4.89\% ($100\% - 95.11\%$) to 21.50\% ($100\% - 78.50\%$).

Similarly, for the MAR metric, \appname consistently requires fewer statements to be inspected across all six fault localization methods compared to the other four data augmentation methods. \appname reduces the average number of statements to be examined to between 74.88\% (for CNN-FL) and 99.23\% (for Dstar) compared to PRAM, equating to reductions of 0.73\% ($100\% - 99.23\%$) to 25.12\% ($100\% - 74.88\%$).

\appname provides significant advantages in fault localization by achieving higher accuracy and reducing inspection effort compared to state-of-the-art data augmentation methods. This enhanced performance demonstrates \appname's potential to streamline the fault localization process effectively.

\begin{tcolorbox}[boxrule=0pt, frame empty, sharp corners, left=0mm, right=0mm]
 \emph{\textbf{Summary for RQ4}: In RQ4, we analyzed the performance of \appname against three data augmentation methods. The results indicated that \appname performs better than Aeneas and Lamont, and slightly better than PRAM.}
\end{tcolorbox}

\section{Discussion}
\label{discussion}

\subsection{Threats to Validity}

\textit{\textbf{The implementation of baselines and our approach.}} Our implementation of the baselines and \appname may potentially contain bugs. As shown in \tablename~\ref{table_dis}, \appname incorporates 3 residual blocks and 3 attention blocks. While this simplified design choice aims to balance complexity and efficiency, it may also reduce the model's capacity to capture more intricate patterns within the data. This could potentially result in underfitting, where the model fails to learn all relevant patterns, especially in cases where the data or the fault localization task requires a deeper model.

\begin{table}[htbp]
  \centering
  \caption{Main Architecture and Parameter of \appname}
    \begin{tabular}{p{16.165em}l}
    \toprule
    \textbf{Main Architecture and Parameter} & \multicolumn{1}{p{13.11em}}{\textbf{Value}} \\
    \midrule
    Number of convolution layers & 13 \\
    Number of residual blocks & 3 \\
    Number of GroupNorm layers & 10 \\
    Number of attention blocks & 3 \\
    \bottomrule
    \end{tabular}%
  \label{table_dis}%
\end{table}%

\textbf{\textit{Dataset-Specific Parameter Choices. }} For the two data augmentation methods, Aeneas and Lamont, both require a parameter that is strongly dependent on the dataset for dimensionality reduction, namely, the number of principal components ($K$). This parameter is inversely proportional to the number of statements in the dataset. In the experiments, $K$ is automatically determined by comparing the number of executed statements across all faulty historical versions in the selected datasets. Therefore, differences in datasets may lead to variations in this parameter, potentially impacting experimental outcomes.

\textit{\textbf{The generalizability.}} Our approach was tested on five representative programs, but its effectiveness on other programs may vary, as no dataset can encompass all fault scenarios. Further experiments on larger programs would be beneficial to confirm the approach's generalizability.

\subsection{Reasons for \appname Is Effective}
The reasons why \appname is more effective than the compared baselines are as follows:
(1) \appname constructs a comprehensive fault semantic context and statistical context from program structure and statistical analysis perspectives. (2) The diffusion model, being a powerful generative approach, ensures effective sample generation without the concern of imbalances between generator and discriminator components, as seen in other models. (3) \appname generates failing test cases to balance the dataset, addressing the class imbalance issue.

\section{Conclusion}\label{conclusion}

In this paper, we propose \appname: a data augmentation approach using a Context-aware and PCA-enhanced diffusion model for fault localization. By employing dynamic program slicing, \appname constructs a fault semantic context that captures the dependencies between faulty statements. A revised PCA is then applied to capture the statistical context. Subsequently, a diffusion model is used to analyze the context matrix and generate synthetic failing test cases creating a class-balanced test suite, significantly enhancing the effectiveness of FL techniques. Our experimental results demonstrate that \appname is effective for FL.

In future work, We will explore the impact of data augmentation for FL in the latent space, rather than original space. Additionally, we will validate our approach's performance on more dataset.


\bibliographystyle{unsrt}  
\bibliography{references}

\end{document}